# CS reconstruction of the speech and musical signals

Trifun Savić, Radoje Albijanić
Faculty of Electrical Engineering
University of Montenegro, Podgorica
trifunsavic@gmail.com, alb.radoje@gmail.com

*Abstract*—The application of Compressive sensing approach to the speech and musical signals is considered in this paper. Compressive sensing (CS) is a new approach to the signal sampling that allows signal reconstruction from a small set of randomly acquired samples. This method is developed for the signals that exhibit the sparsity in a certain domain. Here we have observed two sparsity domains: discrete Fourier and discrete cosine transform domain. Furthermore, two different types of audio signals are analyzed in terms of sparsity and CS performance - musical and speech signals. Comparative analysis of the CS reconstruction using different number of signal samples is performed in the two domains of sparsity. It is shown that the CS can be successfully applied to both, musical and speech signals, but the speech signals are more demanding in terms of the number of observations. Also, our results show that discrete cosine transform domain allows better reconstruction using lower number of observations, compared to the Fourier transform domain, for both types of signals.

## I. INTRODUCTION

Compressive sensing (CS) is a new theory aiming to optimize the signal acquisition process. The main idea behind the CS lies in possibility to reconstruct signal using much lower number of samples than it is determined by Shannon-Nyqiust theorem[1]-[4]. The Shannon-Nyqiust theorem requires samples to be taken in equidistant time intervals, one after another. In CS scenarios, the signal samples should be randomly selected or acquired[1]. By using just small number of signal samples, CS automatically lowers the sampling rate and avoids later signal compression. Also, smaller number of samples means faster signal processing and faster further transmission.

Besides the random selection of the signal samples, CS requires signal to be sparse in a certain transform domain. Sparse signals are characterized by important information condensed in a small number of non-zero coefficients in the sparsity domain[1], [2]. Depending on type of signal, various domains where signals have sparse representation, can be used, e.g., Discrete Fourier Transform - DFT, Discrete Cosine Transform - DCT, Wavelet Transform, etc. Note that the observations are randomly taken from the domain where signal has dense representation. The random selection assures incoherence property to be satisfied. Incoherence and sparsity are two requirements that need to be satisfied in order to apply CS approach.

Generally speaking, the CS can be used for different types of data such as video, images, audio signals, signals used in medical applications (MRI, ECG), etc. In this paper we consider the CS application to the audio and speech signals[5]-[8]. Having in mind that audio signals consist of harmonics, i.e., sinusoid-like components, these signals are good candidates for CS. Namely, a pure sinusoid is a type of signal which has, in the frequency domain, just one non-zero component corresponding to the frequency of the sinusoid. Therefore, sinusoidal signals can be considered as sparse in frequency domain. The sound quality, number of harmonics, played note, could be of different complexity depending on the instrument used to play the sound[9]. Consequently, different number of measurements, used for reconstruction, is required for the signals of different complexity.

Speech signals have more complex nature and therefore they are less sparse in the frequency domain, compared to the pure musical tones. As these signals have short-time stationarity property[10], short length frames should be considered for CS reconstruction. Namely, when observing just a small time interval of the speech signal, it is found that it can be considered as sparse. In this way, successful reconstruction is assured using lower number of available observations.

In this paper we have used different transformation basis for both musical and speech signals, with different number of randomly acquired samples. The quality of reconstructed signals is proved by performing the listening test on the resulting signals, and illustrated by graphical representation of signals in time and frequency domains.

The rest of the paper is organized as follows: In section II theoretical background on CS is presented. The reconstruction of musical and speech signals will be detailed analyzed in section III. In section IV the experimental results are shown, including mean square errors between original and reconstructed signal as well as error versus measurement percentage graph. Conclusion is given in Section V.

## II. THEORETICAL BACKGROUND

Let us introduce discrete signal $f$ in terms of transform domain matrix $\Psi$ and transform domain vectors $x$ as follows:

$$f = \Psi x, \qquad (1)$$

where matrix $\Psi$ is of $N \times N$ ($N$ is the signal length) dimension and $x$ is the signal $f$ transformed into vector in $\Psi$ domain. Vector $x$ is sparse vector, containing $K$ non-zero coefficients, where $K \ll N$. According to the CS scenario, the $M$ samples are randomly taken from the dense vector *f*. In other words, Ψ matrix is subsampled by randomly selecting certain number of columns/rows. Let the matrix $\theta_{M \times N}$ be the sub-matrix of the matrix Ψ, formed by randomly selecting rows/columns of the matrix Ψ. Vector *y*, of $M \times 1$ dimension, is the one containing CS measurements. It is important to emphasize that the number of selected rows/columns *M*, should be much smaller than the signal length, i.e. $M << N$. The vector of measurements can be formed as follows[1]-[3]:

$$y = \theta \Psi x . \quad (2)$$

The equation (2) can also be written as:

$$y = \Omega x . \quad (3)$$

where matrix Ω is of $M \times N$ dimension and it is called CS matrix. As it was mentioned, the length of *y* should be much smaller compared to the length of vector *x*, i.e. $M << N$. By random selection of row/columns of the matrix Ψ, the incoherence property is satisfied and reconstruction can be done from small number of measurements. Equation (3) has larger number of unknowns than the number of equations. Therefore, in order to obtain solution, the optimization algorithms are used. There is a large number of optimization algorithms. Commonly used are greedy algorithms such as Matching Pursuit MP [11], Orthogonal Matching Pursuit OMP[12], [13], Automated single –pass or multi-pass algorithm[14]-[16], but also more complex solutions such as the convex relaxation algorithms. Each algorithm aims at finding the sparsest solution among large number of possible solutions. In this paper, $l_1$-minimization is used. The $l_1$-minimization problem can be defined as:

$$\min_{x} \| x \|_{l_1} \text{ subject to } y = \Omega x , \quad (4)$$

where $l_1$-norm of the signal *x* is sum of the absolute values of the signal samples, and is described with the following relation:

$$\| x \|_{l_1} = \sum_{i=1}^{N} |x_i| . \quad (5)$$

### III. COMPRESSIVE SENSING IN MUSICAL AND SPEECH SIGNALS

The spectral content of a certain musical tone is consisted of several sinusoids located on certain frequencies. The component of the audio signal which is on the lowest frequency is called basic tone (pitch). Others are partial tones or harmonics, and they are multiples of basic tone frequency. Having in mind that sinusoidal signals satisfy sparsity property, we may conclude that the musical tones are convenient for the CS application. These sinusoidal components within the musical signals are called harmonics. The number of harmonics in musical signals certainly affects their sparsity, which consequently affects the number of measurements required for successful reconstruction. The second requirement is called incoherence, and it depends on the samples acquisition procedure. In CS it is necessary that the measurement matrix describing the measurement process is incoherent with the transform matrix.

In this paper, we have used the piano tones and vowels as a test signals. In the piano tone, the sound is generated by vibration of strings. Pressing a key on keyboard causes a padded hammer to hit the string and make them vibrate. In the case of speech, the signal is made by vocal chords, which vibrate as the air flows through them. By losing or tensioning vocal chords we make sounds of different frequencies. It is not necessary for speech sound to be consisted of pitch sound and multiples of its frequency. Speech signal can contain lots of harmonics with non-correlated frequencies. Therefore, speech signal is considered less sparse in observed domains compared to the musical tones.

We have observed two domains, in which signals can be considered as sparse: Discrete Cosine Transform (DCT) and Discrete Fourier Transform (DFT) domains. For the signal *f* of length *N*, the Fourier transform in the discrete form is given by:

$$DFT(k) = \sum_{n=0}^{N-1} f(n) e^{-j2\pi nk/N} , \quad (6)$$

While the DCT can be described using the following relation:

$$DCT(k) = c(k) \sum_{n=0}^{N-1} f(n) \cos \frac{(2n+1)k\pi}{2N} , \quad (7)$$

Where coefficients *c(k)* are defined as:

$$c(k) = \begin{cases} \sqrt{1/N}, & k = 0 \\ \sqrt{2/N}, & k = 1, ..., N-1 \end{cases} \quad (8)$$

By using different type of transform basis, we got different results for the same number of acquired measurements. Comparative analysis regarding these two domains will be given in the following.

Having suitable sparsity domain, the measurements of the signal are taken from the domain where signal is dense. Thus, the measurements of the audio and speech signals are taken randomly from the time domain. The CS measurements matrix *Ω* is made from the transform domain matrix, by randomly choosing *M* rows of the basis matrix *Ψ* (DFT or DCT transform matrix).

$$\psi = DFT \ \vee \ \psi = DCT . \quad (9)$$

Random selection is performed by random permutation of the vector *q*, which contains positions from 1 to *N*. Taking just first *M* coefficients form the vector *q*, we have in fact defined *M* random rows to be selected from the matrix ψ. Mathematically, this can be described as:

$$q = randperm(N) , \quad (10)$$
$$B = inv(\Psi) , \quad (11)$$
$$\Omega = B(q(1:M),:) . \quad (12)$$

Now, the vector of measurements is given by:

$$y = \Omega x = B(q(1:M),:)x \quad (13)$$

The previous relation can be solved by using $l_1$ norm minimization, according to is relation (4).

### IV. EXPERIMENTAL RESULTS

In this section we have observed two types of signals and therefore we have divided our experiment into two parts: 1) the reconstruction analysis for audio/musical signals, and 2) analysis for the speech signals. For both signals, the two transform domains are observed: DFT and DCT domain. The experiments are repeated several times, with different number of available signal samples. Mean square error (MSE) is calculated after each reconstruction, and MSE versus number of measurements graphs are displayed.

*Example 1: Audio signal*

For experiments with the audio signals, the piano tone is chosen with the length of 3000 samples. The reconstruction performance is analyzed for different number of measurements: 20% - 90% samples available. Also, both the DFT and the DCT transform basis are observed as sparsity domains. Figure 1 shows the results of reconstruction using 30% and 50% samples, for both basis. It can been concluded that, by using the DCT as sparsity domain only 30% of the available samples are enough for successful reconstruction. When using the DFT as a sparse basis, the number of measurements needs to larger such as 50% of the total signal length.

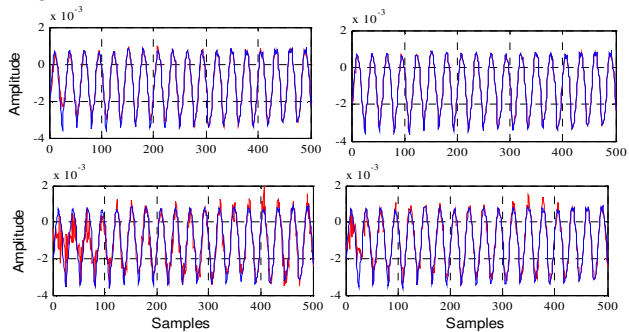

Figure 1.Original piano signal (blue), reconstructed (red), using 30% DCT (upper left graph), 50% DCT (upper right graph), 30% DFT (lower left graph), 50% DFT(lower right graph), time domain

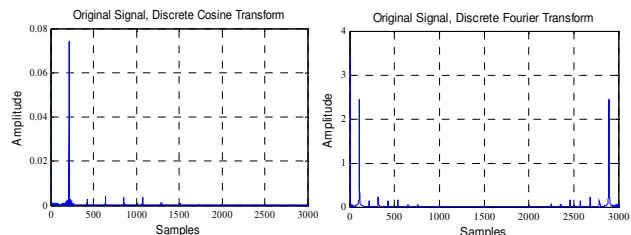

Figure 2.Original piano signal in DCT (left graph), in DFT (right graph) domain

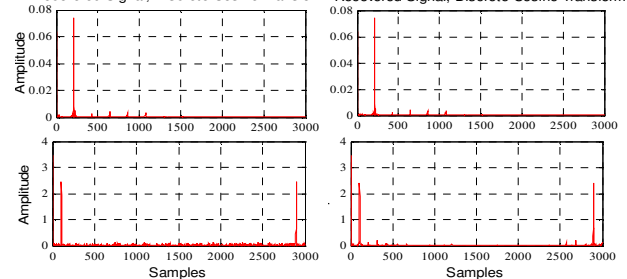

Figure 3.Reconstructed piano signal using 30% DCT (upper left graph), 50% DCT (upper right graph), 30%DFT (lower left graph), 50% DFT(lower right graph), frequency domain

Figure 2 shows the original signal in the frequency domain (DCT and DFT), while the reconstructed signal in DCT and DFT domains are shown in Figure 3 (with 30% and 50% samples used for the reconstruction). As a more relevant indicator of reconstruction quality, in Figure 4 we ploted the mean square error (MSE) between reconstructed piano signal using DCT and DFT as transform domains, for different number of measurements. Again, it is shown that the MSE is much lower in the case when DCT is used as a sparse basis.

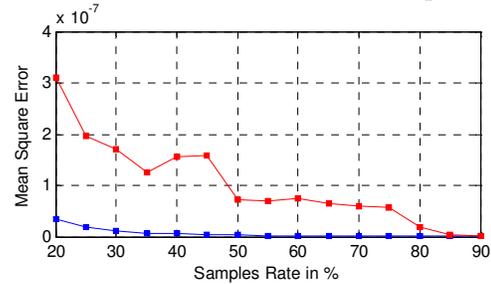

Figure 4. MSE vs number of measurements for the piano signal; DCT (blue), and DFT (red)

*Example 2: Speech signal*

In this example, speech signals with length of 3000 samples are observed. Experiments are repeated several times, using different number of samples as CS measurements: from 20% to 90% of the original number of samples. Figure 5 represents time domains of the original and reconstructed speech signal (as a represent of speech signal we illustrated the vowel E). The graphs are for the 50% and 70% of samples used as measurements and with the DCT and the DFT domains. The DCT gives satisfactory reconstruction using 50% of samples, while the DFT requires 70% of samples for the successful reconstruction.

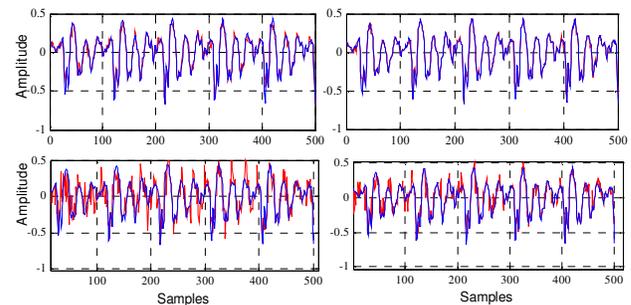

Figure 5.Original speech signal (blue), reconstructed (red), using 50% DCT (upper left graph), 70% DCT (upper right graph), 50%DFT (lower left graph), 70% DFT(lower right graph), time domain

Therefore, we can again conclude that DCT domain is more applicable as sparse domain, for both types of signals. Figures 6 and 7 show frequency domains of original and reconstructed signals.

The graph representing the MSE in terms of the number of measurements graph, is presented in Figure 8. The DCT and the DFT domains are used in reconstruction and MSE is calculated for both domains. As we can see from experimental results, DCT transform basis gives smaller MSE for the same number of used measurements, compared to the DFT.

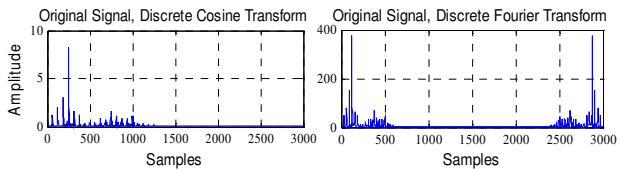

Figure 6. Original speech signal using DCT (left graph), using DFT (right graph), frequency domain

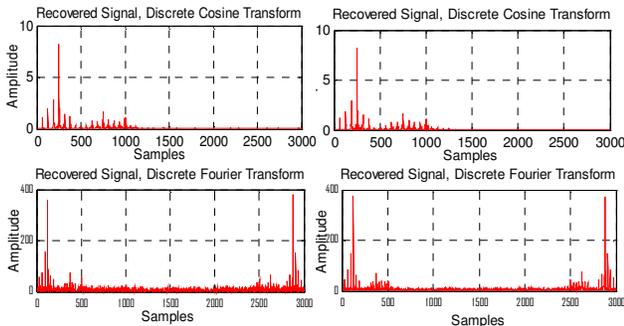

Figure 7. Reconstructed speech signal using 50% DCT (upper left graph), 70% DCT (upper right graph), 50%DFT (lower left graph), 70% DFT (lower right graph), frequency domain

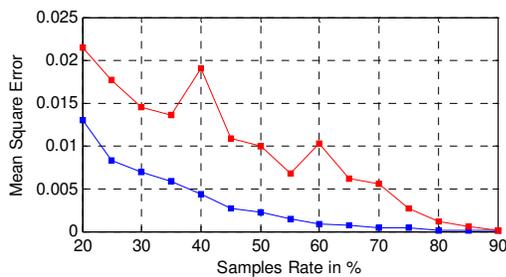

Figure 8. MSE using DCT (blue), and DFT (red), speech signal

## V. CONCLUSION

In this paper we analyzed the Compressive Sensing application to the speech and audio signals. Different transform domains are tested, in order to choose suitable basis for successful reconstruction from small set of available samples. It is shown that the DCT domain requires smaller number of samples compared to the DFT, for both types of signals. Moreover, when observing the MSE for audio signal, it is shown that reconstruction with 20% of measurements and the DCT as transform basis, gives the same error as 80% of measurements and the DFT as transform basis. Further, the audio signals are less complex compared to the speech signals and therefore much smaller number of samples is required for reconstruction of audio signals. Audio signals can be reconstructed with 30% of samples, while for the speech signal this number is slightly larger – 50%.

## VI. ACKNOWLEDGEMENT

The authors are thankful to Professors and assistants within the Laboratory for Multimedia Signals and Systems, at the University of Montenegro, for providing the ideas, codes, literature and results developed for the project CS-ICT (funded by the Montenegrin Ministry of Science).